\documentclass[fleqn,10pt]{wlscirep}

\title{L\'evy like patterns in the small-scale movements of marsupials in an unfamiliar and risky environment}

\author[1] {B. R\'{\i}os-Uzeda}
\author[2,*]{E. Brigatti}
\author[1]{M. V. Vieira}
\affil[1]{Laborat\'orio de Vertebrados, Instituto de Biologia, Universidade Federal do Rio de Janeiro. Caixa Postal 68020, 21941-590, Rio de Janeiro, RJ, Brasil.}
\affil[2]{Instituto de F\'{\i}sica, Universidade Federal do Rio de Janeiro, 
Av. Athos da Silveira Ramos, 149,
Cidade Universit\'aria, 21941-972, Rio de Janeiro, RJ, Brasil.}
\affil[*]{edgardo@if.ufrj.br}

\begin{abstract} 

We investigate the movement patterns of three different
Neotropical marsupials in an unfamiliar and risky environment.
Animals are released in a matrix 
from which they try to reach a patch of forest.
Their movements, performed on a small spacial scale, 
are best approximated by L\'evy flights.
Patterns of oriented and non-oriented individuals - with forest patches within 
or beyond their perceptual range - differ only slightly in the value 
of their exponents. 
These facts 
suggest that, for these species,
the appearance of L\'evy flights is 
the product of animals innate 
behaviour 
that emerges spontaneously, as a neutral characteristic proper of a 
default movement mode for alerted animals.
\end{abstract} 

\begin{document} 
\flushbottom
\maketitle

\section*{Introduction}

Loss and fragmentation of natural habitats affects ecological processes such as species interactions, trophic dynamics and dispersal processes, and can cause declines in population density, changes in species richness and  in community composition \cite{2,3,4}.
Habitat loss and fragmentation also affects movements,
making it difficult for an animal to move in unfamiliar enviroments, such as the matrix between fragments 
and, to reach new suitable habitats \cite{7,8}. 
In the Neotropics, marsupials are an example of small non-flying vertebrates affected by fragmentation \cite{4}. 
Most species of Neotropical marsupials are forest species, inhabit 
the forest and use the matrix just to move between patches. 
Although predation risk is likely higher in the matrix, within the forest they are also susceptible
to be predated by a wide range of animals, e.g. snakes or raptors  \cite{10}. For this reason, in both forest and matrix specific patterns of movement are important to minimise predation risk, to increase survival probabilities  \cite{12}, and the success in reaching a forest patch.

In movement ecology, the description  of animal movement 
can be based on the characterisation of different stages
that can be related to three questions: when to move, where to move, and how to move  \cite{13,14}. 
When to move involves factors related with evaluation of food, shelter, or reproduction availability. 
Where to move is generally determined by
the previous knowledge of the environment, perceptual capacity, and ability of orientation.
How to move, is related to locomotion ability, and 
to numerous internal and external factors that can provide assistance to movements, improving the chance of reaching the final location.  

A first step in the study of animal movement patterns is the
characterisation of  individual paths measuring 
their turning angles (the difference between previous and current direction of motion), 
and their step lengths (the distance between the turning points)  \cite{15,16,17,18,19}. 
This type of analysis has been frequently carried out for trajectories in two dimensions 
(see for example  \cite{20,22}), 
and sometimes also in three dimensions  \cite{23}. 
A more comprehensive description of movement patterns 
should also consider the temporal dynamics of these 
trajectories, characterising the movements by means of 
the spatial and temporal distributions of animal positions.
This approach, in analogy with classical models 
of diffusion of particles, allows to explore the mathematical and 
statistical properties of these distributions, and to connect them 
with the ecological processes underlying these movements \cite{18,24,25,26}.

Historically, a first tentative description of  
animal movement patterns introduced the use of 
a simple random motion, generally described 
by means of a Brownian motion. 
In a second moment, a new approach 
was introduced with the use of
correlated random walks, 
a probabilistic model where movements present some type of memory of the 
previous motion \cite{15}. 
Correlated random walks  were introduced with the aim of 
teasing apart searching and foraging from random movement behaviour, 
 and they can be considered a milestone in the development of movement ecology discipline \cite{14}. 
Later, an alternative description of animal movements was proposed with the introduction of L\'evy processes \cite{levy1}.  
This approach originated in the field of statistical mechanics, 
and nowadays presents large application in physics and 
natural sciences \cite{35}. 
It presents characteristic trajectories
marked by abrupt long jumps which connects 
clusters of frequent short displacements,
generating a characteristic scale-invariant structure 
and a possible super-diffusive behaviour \cite{26,49}.
Since the end of the 90's 
this process has attracted the attention of the ecologists
for being considered a result of optimal foraging theory \cite{16,30}, 
and a large amount of studies, connected with its use for describing
searching and foraging movements, has flourished since then
(e.g. Viswanathan et al. \cite{16,30,37,38,39,57}, Bartumeus et al. \cite{26,31,40,31}, Bartumeus \cite{42,43}, Edwards \cite{17}, Edwards et al. \cite{44,45}, and Reynolds \cite{22,25,46,47,48}).

Controversy around this approach emerged, 
questioning the results of many early studies, 
generally because of the use of inappropriate statistical techniques \cite{17,44,45}.
However, in the last ten years, 
many studies 
confirmed that L\'evy flights are a good approximation 
to the patterns of movement of different species,
either in the case of foraging animals looking for scarce preys,
or emerging from a general innate behaviour \cite{22,25}.
However, cases of L\'evy processes in nature involved movements of individual organisms over large areas, during many days or periods of activity. Also, animals were engaged in foraging activities, whether in a familiar environment or not. 

The implementation of the three previous movement models (Brownian motion, CRW, and L\'evy processes) in terms of discrete models is simple.
The common idea 
is to assume that the 
movement of an animal consists in a discrete 
series of steps separated by events of reorientation 
\cite{17,30,31}. 
The step-length $l$ 
is selected from a distribution $P(l)$ and
its direction, defined as the turning angle in relation to the 
the direction of the previous movement, is picked 
up from a distribution $P(\theta)$.
The Brownian random walk is characterized by a $P(l)$ with a finite variance.
Generally it corresponds to a Gaussian but even a fixed step length is 
a possible choice. 
The direction of every step is isotropic, with $P(\theta)$ represented by a uniform random distribution.  
The correlated random walk presents $P(l)$ with a finite variance. However, the directions of the movements are not isotropic, introducing a correlation between the directions of subsequent movements. In fact, the  persistence in the direction 
(the degree of correlation of the random walk) is controlled by the shape of the turning angles distribution \cite{15,33}. Generally, this effect can be obtained using a $P(\theta)$ following a wrapped Gaussian distribution.

In the case of the L\'evy processes, usually called L\'evy flights,  step-lengths have a probability distribution that is heavy-tailed: $P(l) \propto l^{-\mu}$, with $1<\mu\le3$, and the directions are isotropic \cite{30,32}. 
Note that $\mu>3$ generates trajectories with large-scale 
properties typical of a Brownian motion, 
whereas, approaching $\mu=1$, ballistic trajectories are recovered.
Le\'vy flights have been suggested to be the best solution
for solving the problem of random search \cite{16,16b}.
In particular, Viswanathan et al. \cite{30} proved that  
in environments where the resources are homogeneous and scarce, 
if search targets are not depleted or reject once visited, but can instead 
be profitably revisited,
the optimal $\mu$ is 2.
This result generalises to  $\mu \approx 2$ in fragmented landscapes \cite{104}.\\




The main purpose of this study is to characterise the movement patterns of
three Neotropical marsupials released in a matrix of an unfamiliar habitat.
The three species represent the range of body sizes of Neotropical marsupials that are generalist, mostly omnivores, and are capable of climbing but mostly use the forest floor or understory. 
The matrix in the study areas was  composed of pastures, which represents an unfavourable and risky habitat for these animals. 
For this reason, we can consider that  it
induces animals to keep alert of potential predations and to seek for a safe place,
which, in our experimental set-up, is a patch of forest at a given distance.
This is an important difference in relation to the majority of the existent literature,
which has reported results related to movements of animals
searching for food. 
Simulations suggest that predation risk may favour smaller values of the $\mu$ exponent of 
L\'evy walks \cite{Shimada,103}, but may also favour large values of $\mu$ depending on the foraging strategy of the predator \cite{Shimada}. 
Another relevant novelty is that our study is developed on spatial 
and temporal scales smaller than
the ones usually considered in previous studies with 
larger animals.
The trajectories of the small marsupials studied here were performed within a small region of the matrix, and represent only part of one cycle of activity of an individual (less than one day). 

Finally, 
two distinct situations were considered: 
animals that are able to orient themselves in the direction of the patch,
and animals that are not able to orient themselves. 
In this way, we are able to unfold a rich comparison between the movements of three species representing a range of body sizes, both for oriented and non-oriented animals.

\section*{Material and Methods}

\subsection*{Study site and field methods}

The dataset is the result of a series of studies performed from 2007 to 2010 to determine the perceptual ranges and movement behavior of marsupials in an Atlantic Forest landscape in Rio de Janeiro state, Brazil. 
The field work was conducted in the Guapi-Macacu river basin ($22\,^{\circ}25'$ S and $42\,^{\circ}44'$ W), which is part of the municipality of Guapimirim and Cachoeiras de Macacu.
The climate is mild humid-mesotermic \cite{Nimer89}, and vegetation of the region is classified as dense evergreen forest (``Ombr\'ofila Densa"; \cite{IBGE}). Vegetation of the fragments is disturbed to various degrees, with a relatively open understory and canopy, and is characterized by the presence of palms ({\it Astrocaryum aculeatissimum}), {\it Cecropia sp.}, and lianas \cite{Finotti}. 
The region of the Guapi-Macacu River basin has a long history of human occupation, and currently the landscape is characterized mainly by small forest fragments ($< 200$ ha) structurally isolated by a matrix of urban areas, pastures, plantations, and paved roads \cite{Cabral,Pedreira,Vieira09}. 
The Guapi-Macacu river basin has ca. 45\% of forest cover, part of it old-growth forest at the base and on the slopes of sierras, along the northern portion of the river basin \cite{Fidalgo}. The matrix where forest fragments are inserted is heterogeneous, with different plantations and pasture systems \cite{Cabral,Pedreira,Uzeda}.

Perceptual ranges were determined by releasing individuals in the matrix at different distances (30, 50, 100 and 200m) from an unfamiliar forest patch. The three species studied are {\it Didelphis aurita, Marmosa paraguayana} and {\it Philander frenatus} and they represent the range of body sizes of Neotropical marsupials (mean body mass of individuals released: 850g in {\it D. aurita}, 395g in {\it P. frenatus}, and 115g in {\it M. paraguayana}).
In addition to the tracking of trajectories, the experimental set-up allowed 
to estimate the ability  of animals to orient themselves in the direction
of a forest patch 
and to classify the data on the bases of this characteristic.

As in \cite{74}, it was assumed that animals, being released in an open and unfamiliar habitat, would
look for refuge which would corresponded to the nearest patch of forest.
All animals were released just once to avoid possible cumulative experiences,
and  the site of release was distant, at least, 1000 m from the point of capture to reduce the possibility of familiarity with the region. 

\begin{figure}
\centering
\includegraphics[width=0.7\textwidth, angle=0]{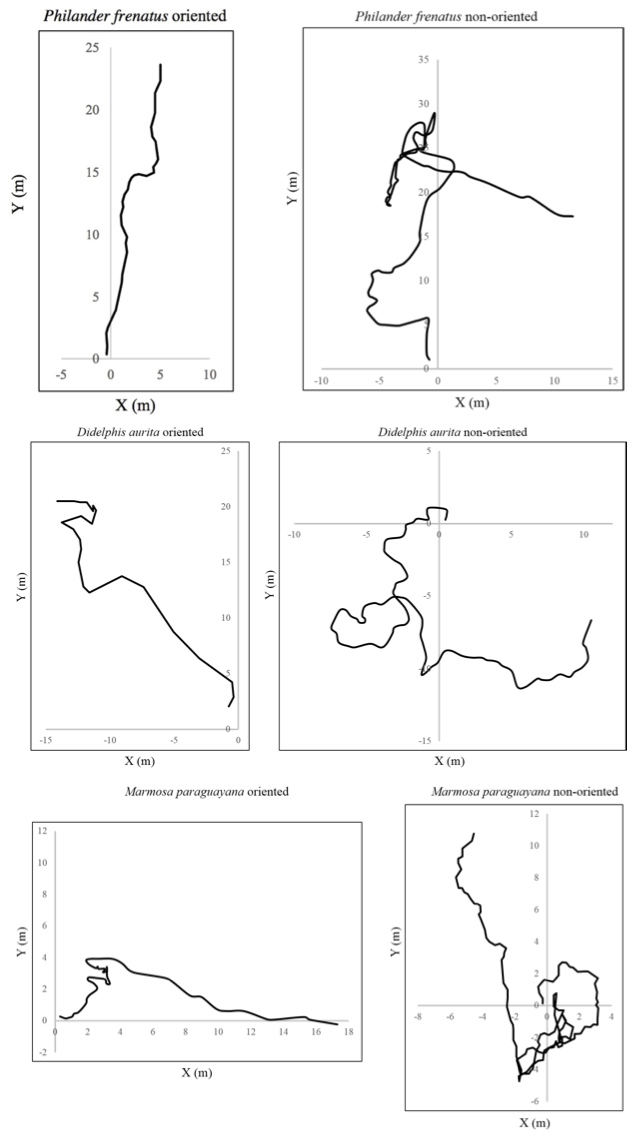}
\caption{ {\bf Experimental trajectories.}
Examples of the trajectories of oriented and non-oriented individuals of the three species of marsupials. 
Oriented individuals of {\it P. frenatus} and {\it M. paraguayana} were released at 100 m from the nearest forest fragment, 200 m in {\it D. aurita}; non-oriented {\it P. frenatus} and {\it M. paraguayana} were released at 200 m from the nearest forest fragment, 300 m in {\it D. aurita.}
Note that, at the scale of these figures, the body sizes of these animals are of the order of a point.  
The data-set variability, even inside each class, is very hight. 
}
\label{fig_trajectory}
\end{figure}

Individuals were equipped with a spool-and-line device, that consists of bobbinless cocoons of nylon thread wrapped in a polyvinyl chloride film \cite{72,Boonstra,Loretto}. Each device weighs $4.8$ $g$ and the thread 
is $480$ $m$ long \cite{Delciello}. Devices were attached to the fur between the shoulders of each individual by using an ester-cyanoacrylate-based glue (Henkel Loctite Adesivos Ltda., Manaus, Brazil). 
The thread released by the device allowed mapping the animals' path (see Figure \ref{fig_trajectory}). The direction of the path was measured aligning a compass with the direction of the line to the next point of change of direction greater than five degrees (as in \cite{74}, but without the rediscretisation to 10 degrees in \cite{75,75b,prevedello}), and the linear distance between points was measured with a tapeline.
The experiment ended when the animal reached the target or it 
went out of the matrix \cite{75}. 
Paths were followed up to 170m, to allow more individuals to be tracked. 
Occasionally the experiment ended before 170m of path, when the animal 
reached the target \cite{prevedello}. 
Trapping and handling conformed to guidelines sanctioned by the American Society of Mammalogists \cite{Sikes2016}. This study was approved by IBAMA/MMA (Authorization numbers 87/05-RJ, 099/06-RJ, 13861-1, 13861-2, 16703).

Trajectories were classified as belonging to an oriented 
or a non-oriented animal on the basis of the estimated
perceptual range of these species,
defined as the ability to perceive 
a forest patch at a given distance from the 
release point. 
The estimation of the perceptual range was obtained in \cite{74,75b} 
evaluating the mean vector of orientation of the first 20m of path 
which was used to determine if individuals released at a certain distance were significantly oriented towards the nearest forest patch. Distances to a forest patch beyond perceptual range where the ones with non-significant concentration of angles in the direction of the closest forest patch.
Perceptual ranges were estimated as $100$ $m$
for {\it M. paraguaiana} and {\it P. frenatus}, and $200$ $m$ 
for {\it D. aurita} \cite{74}. 
Based on these results, we considered as oriented the animals
released at a distance smaller or equal to the perceptual range,
and non-oriented the other cases.


%

\subsection*{Statistical modeling and parameter estimation}

The principal aim of our analysis is to evaluate the distribution of 
the step-lengths $l$,
defined as the distance between two successive turning points.
If the measured trajectories are directly used to evaluate $l$,
the turning points are simply selected as all the points with an angle 
larger than a minimal angle $\theta_{min}$. 
For our data set it would be natural to choose $\theta_{min}=5$.  
However, this ad-hoc discretization of the movement patterns would be arbitrary
and the estimated distribution $P(l)$ would strongly depend on this choice \cite{23}.
We overcame this problem with a recent approach introduced in \cite{32},
which analyse not the trajectories, but their projections.
In analogy with that procedure, a trajectory in 2D is projected on one 
axis, the $x$-axis for example. 
A projected step $l_x$ is defined  as 
the distance between two inversions in the direction of the movement of the
projected trajectory. 
 Similarly, we obtain the step-lengths along the $y$-axis 
 ($l_y$).
It was analytically proven that if the distribution of steps in the original 2D path followed a power law,
the distribution of the projected paths preserves the 
same power-law relationship \cite{32}. 
In the case of an exponential distribution, the projected 
data do not preserve the same functional form, but they maintain the 
general exponentially decaying behaviour.
Following this method we were able to investigate the shape of the 
$P(l)$ distribution using a segmentation procedure that
does not depend on any arbitrary choice.
Moreover, any correlation between turning angles in the original dataset 
would be eliminated in the projected data.
As pointed out in \cite{32,23}, 
the operation of projection, however, causes the proliferation of spurious 
data corresponding to projected steps 
smaller than the minimum 2D step-length present in the original dataset.
These data must be excluded from the final analysis.
Moreover, as the last measured step-length can have a length influenced 
by the ending of the spool, 
they were eliminated from the data set. 
It follows that also rare events, with a 
unique straight trajectory were excluded. 
In this way, pure ballistic movements are 
ruled out.
Finally, after having checked that the two distributions of $P(l_x)$ and $P(l_y)$, 
as expected, do not show any discrepancy, we combined in $P(l)$ the data coming from the two projections. 

Once obtained $P(l)$, we face the problem of estimating
the best model that describes it. 
The characterisation of a power-law behaviour could be
obtained
using a simple Pareto distribution:
$f(x)=(\mu-1)a^{\mu-1}x^{-\mu}$, for $x\ge a$. 
However, for our dataset this distribution should be ruled out 
because surely important truncation effects are present
and a limit to the maximum $l$ value must be taken into account.
A realistic animal movement is characterised by physiological
and circadian rhythms that cause direct limits on the 
maximum step length.
Second, and more important, 
inevitable truncation effects come from the measurement 
process, which cannot 
measure step-lengths larger than the length of the spool.
For these reasons, it is more appropriate the use of a 
truncated Pareto distribution, defined as:
$g(x)=(1-\mu)/(b^{1-\mu}-a^{1-\mu})x^{-\mu}$, for $a \le x\le b$.
Although for reasonable high values of 
the $\mu$ exponent the use of the Pareto or truncated Pareto distribution 
can lead to comparable results,
for small values of $\mu$ (typically $\mu<2$) the difference in the estimation is substantial, as can be proven by numerical tests, and the use of the Pareto truncated distribution is essential. 

The best estimation of the parameters of this distribution was 
obtained using the Maximum Likelihood Estimation (MLE),
which is more
reliable compared to classical least squares methods (see, for example, \cite{89,90} 
for a detailed discussion).
For a truncated Pareto distribution, the estimation of the $\mu$ parameter  is given by the numerical solution of the equation \cite{90,91}: 
\begin{equation}
\frac{1}{n}\sum_{i=1}^n ln(l_i)= \frac{1}{\mu-1}+\frac{b^{1-\mu}ln(b)-a^{1-\mu}ln(a)}{b^{1-\mu}-a^{1-\mu}}
\label{eq:MLE}
\end{equation}
where $a=min(l_i)$ and $b=max(l_i)$. 

Unfortunately, in contrast to the Pareto distribution, 
 there is no simple tests to determine whether a truncated Pareto model
generates an appropriate goodness of fit \cite{91}.
For this reason, the estimation must be supplemented by a graphical 
check of the data. 
The inspection was realised comparing the best fit model
 to the frequency of the step-lengths of the projected 
 movements using a log-binned histogram. Survival functions
were not introduced because, for small values of $\mu$, it is
not possible to make
 a clear comparison \cite{32}.
 
A model selection approach  was used to compare models based on the whole dataset, considering all step-lengths. To quantify the evidence supporting each model
we used the Akaike information criterion ($AIC$), which compares models likelihoods penalizing models with more parameters \cite{Anderson}.
The $AIC$ is calculated as follows: $AIC=2K-2L$,
where $L$ is the maximum log-likelihood and $K$ is the number of parameters of the
model. The maximum log-likelihood was estimated following Edwards \cite{45}.
The model with the lowest Akaike information is the best supported model.
In addition to the Pareto and the truncated Pareto distribution,  we consider also
the exponential distribution: $h(x)=\lambda\exp(\lambda a) \exp(-\lambda x)$ for $x\ge a$.

The model selection approach was also used to compare support between exponential and truncated Pareto for different subsets of the data. 
This approach can be useful for detecting the 
possible existence of some typical scale, corresponding to the occurence of an 
exponential truncation. 
This analysis was obtained 
performing a sequential maximum likelihood estimation of model parameters and  a model selection 
on different subsets of our data.  
A given subset $n$ was defined from the previous $n-1$ set by subtracting its smallest value. 
Looking at the $AIC$ weights for each subset, plotted as a function of its smallest value, this procedure can tell us if a critical step value exists where there is more support for an exponential rather than a truncated Pareto distribution. \\

After having analysed the projected data, we also 
inspected the behaviour of the original empirical data, 
estimating the corresponding distributions $P(l)$ and $P(\theta)$,
and
we 
implement a simulation 
based on a discrete walk, for synthetically reproducing the
empirical data-set.
Our aim is to test the robustness of our analysis
and to clarify the relationships linking the original and the projected data. 
For this part of the study we focus our attention on the data of oriented individuals of {\it Philander frenatus} because
these data are abundant and present well behaved distributions.\\

Most of the data generated and analysed during the current study are included 
in this article and its Supplementary Information. The rest of the raw data 
are available from the corresponding authors upon request.


\section*{Results}

\begin{figure}
\centering
\includegraphics[width=0.5\textwidth, angle=0]{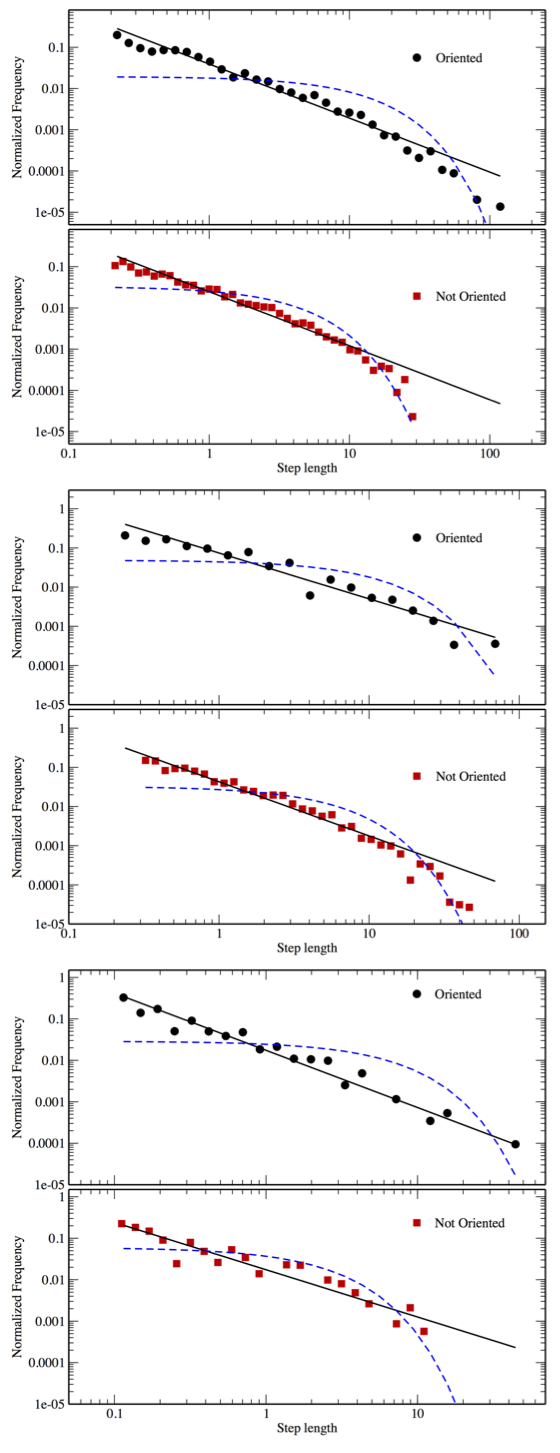}
\caption{{\bf Step-lengths distributions.}
Log-binned data with the best
estimated Pareto-truncated (continuous line) and exponential distribution (dashed line).
Step-lengths are expressed in meters.
From top to bottom: {\it Philander frenatus, Didelphis aurita,  Marmosa paraguayana}.
Black dots are data from oriented animals, red squares from non-oriented ones.
 }
\label{fig:fitting}
\end{figure}

The 
Pareto-truncated distribution always
presents the lowest $AIC$ values considering the whole dataset
(see Table \ref{table0}).
Support for the Pareto-truncated is much higher than the
competing distributions, 
as indicated by the $\Delta AIC$ values, which, in the worst case,
is as hight as 28,  
and the Akaike weights, which suggest that the distributions cannot easily be 
mistaken for Pareto or exponential distributions (see Supplementary Information for further details on the model selection).
A visual inspection is sufficient to persuade that
the fitting of the exponential to our dataset is very poor (see Figure \ref{fig:fitting}).
Table \ref{table} resumes the results of this analysis for the 
different species, distinguishing between oriented and non-oriented animals. \\

\begin{table}
    \centering
{\small
    \begin{tabular}{|c|c|c|c|c|}  
\hline
Species &   & Pareto-truncated & Pareto & Exponential  \\
\hline
\hline
\it Philander frenatus & Oriented  & 2529  & 2727 &  3522  \\
\hline
 & Non-oriented  &  4458 & 4951 &  5246  \\
\hline 
\hline
\hline
\it Didelphis aurita & Oriented & 992 & 1080 & 1236   \\
\hline
 & Non-oriented & 3249 & 3662 & 3755    \\
\hline
\hline
\hline
\it Marmosa paraguayana & Oriented  & 227 & 255 & 465 \\
\hline
 & Non-oriented  & 179 & 215 & 209   \\
\hline
\hline
\end{tabular}
}
\caption{ The Akaike information criterion values ($AIC$) for the Pareto truncated, Pareto and exponential
distributions. Further results are given in the Supplementary Information.
}
\label{table0}
\end{table}

\begin{table}
    \centering
{\small
    \begin{tabular}{|c|c|c|c|c|c|c|}  
\hline
Species &   &  Observations &Sample Size & $\mu$ & $a$ (m) & $b$ (m) \\
\hline
\hline
\it Philander frenatus & Oriented  & 42 & 541   &  $1.31\pm0.03 $  & 0.20 & 128.76\\
\hline
 & Non-oriented  &  57 & 1161 &  $1.38\pm0.02 $  & 0.22 & 59.46 \\
\hline 
\hline
\hline
\it Didelphis aurita & Oriented & 19 & 182  & $1.17\pm0.05 $   &  0.23 & 79.63 \\
\hline
 & Non-oriented & 41 & 875 & $1.38\pm0.02  $  &  0.22 & 48.82 \\
\hline
\hline
\hline
\it Marmosa paraguayana & Oriented  & 6 & 88 & $1.38\pm0.07 $  & 0.10 & 40.03 \\
\hline
 & Non-oriented  & 3 & 65 & $1.14\pm0.09$   & 0.10 & 11.20 \\
\hline
\hline
\end{tabular}
}
\caption{ Values of $\mu$ for the Pareto truncated distribution for the three species of marsupials. 
Observations indicate the number of trajectories analysed.
Each trajectory is the whole track of a single individual.
Sample size corresponds to the number of regressed data 
after eliminating the spurious $l$ smaller 
than the minimal step of the original (not projected) data.
The values of parameters $a$ and $b$ are expressed in meters.
}
\label{table}
\end{table}

\begin{figure}
\centering
\includegraphics[width=0.9\textwidth, angle=0]{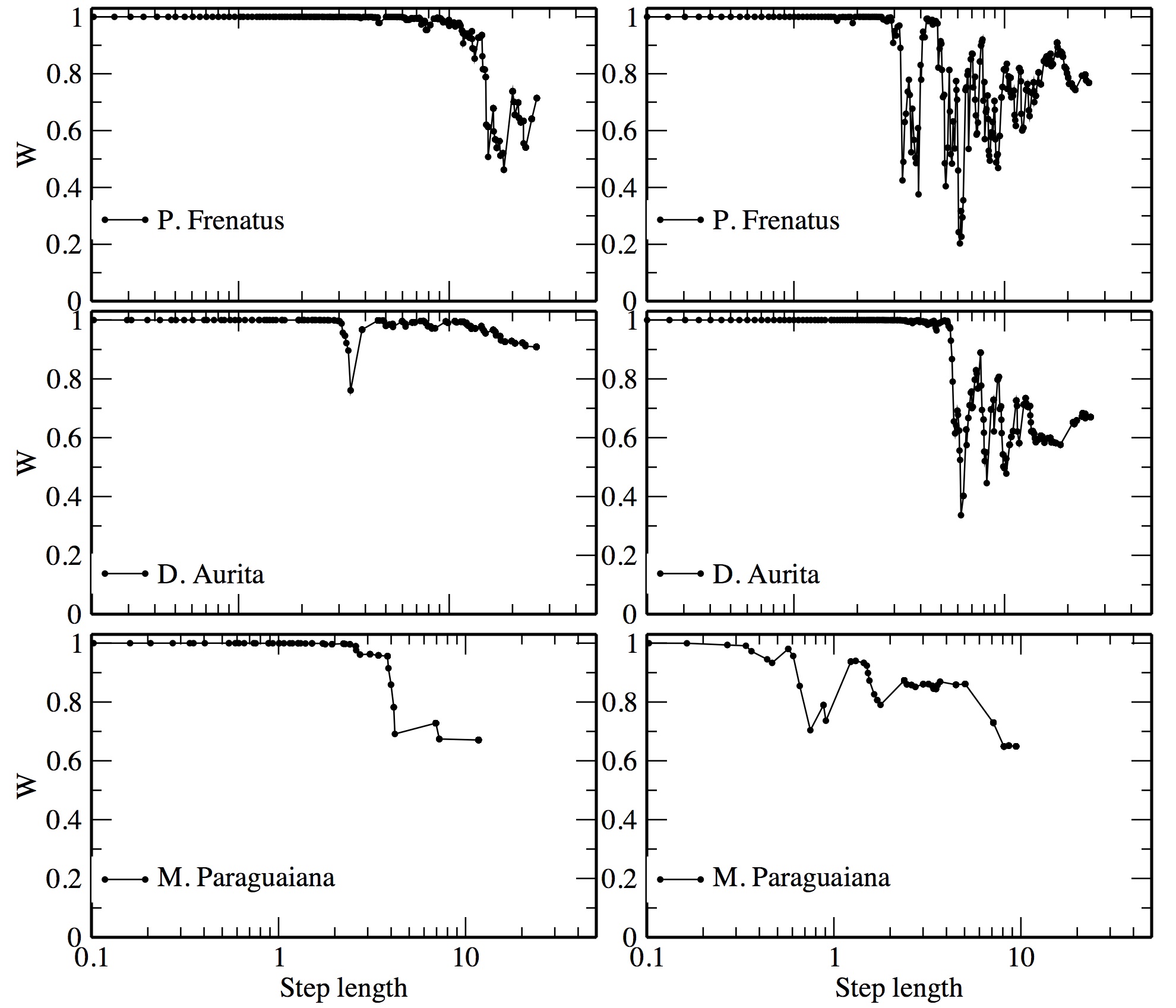}
\caption{{\bf The sequential maximum likelihood estimation}.
Weight of evidence (Akaike weights) favouring Pareto-truncated compared to exponential distribution along different subsets of the step lengths, for oriented (left) and non-oriented (right) individuals. Akaike weights were estimated by sequential maximum likelihood, and displayed as a function of the smallest step length in each subset.
A  weight of one corresponds to the maximum weight of evidence in favor of   the Pareto-truncated distribution.
 }
\label{fig:scale}
\end{figure}

Based on the sequential maximum likelihood estimation on different subsets of our data, 
only non-oriented {\it D. Aurita} and {\it P. Frenatus} present a large step value for which the weight of evidence in favor of the exponential model grows, but rarely passing the critical 0.5 value (see Figure \ref{fig:scale}). 
In all the other cases 
the power-law model is 
the most probable one at all scales, even if there is a general deterioration in the evidence. 
Therefore, the inference of power law behaviour is robust, and there is no evidence of best exponential fit on subsets of the data. \\
 
\begin{figure}[h]
\centering
\includegraphics[width=0.75\textwidth, angle=0]{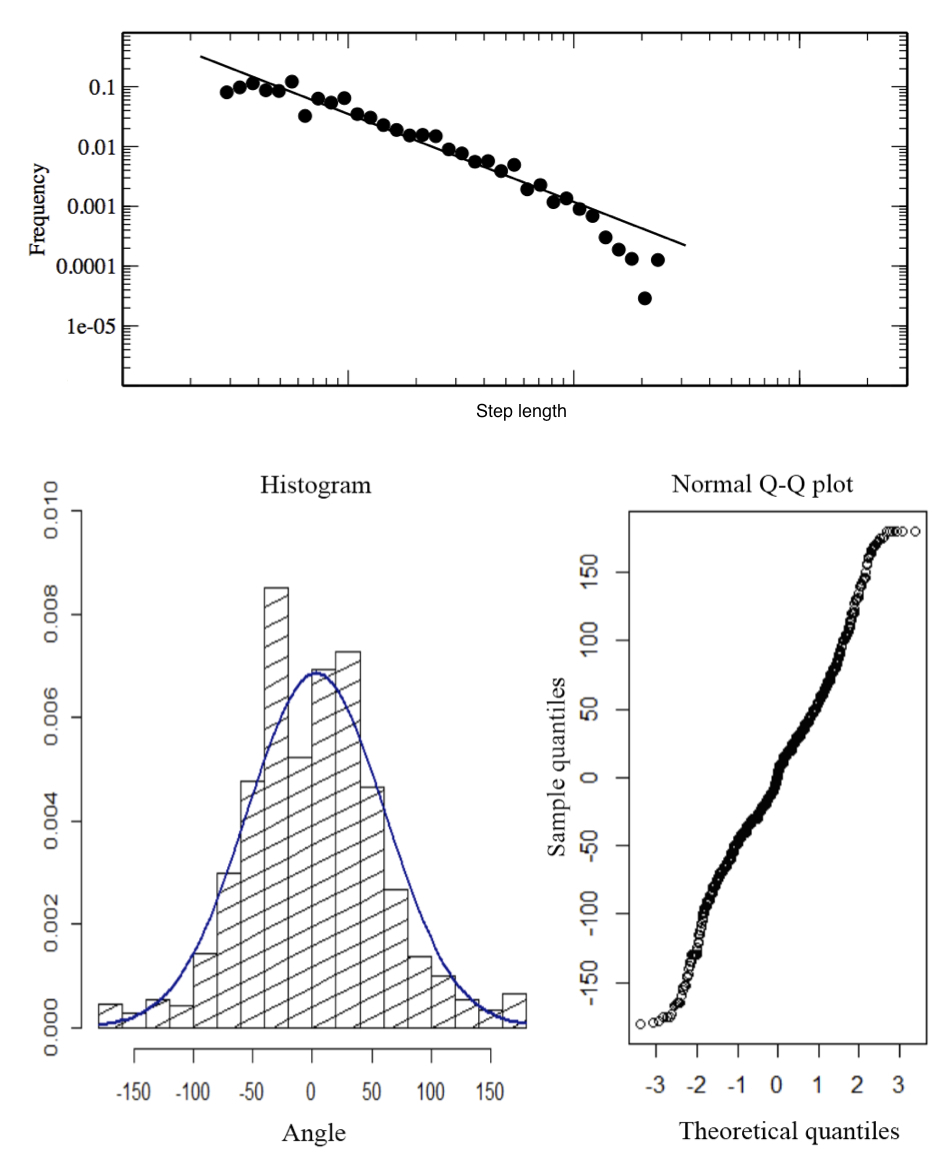}
\caption{{\bf Original empirical data.}
Top: 
Distribution of step lenghts of the original empirical data 
for oriented {\it Philander frenatus}. 
Step lengths are log-binned. 
The continuous line represents the best
estimated Pareto-truncated distribution.
Bottom: Turning angle distribution of the same data. 
Normality of the data can be appreciate looking at the q-q-plot. 
 }
\label{fig_empiric}
\end{figure}

For the analysis of the original empirical data, step-lengths ($l_o$) 
are defined as the distance between two turning points.
This is equivalent to set the threshold of the turning angles to $5\,^{\circ}$.
The inspection of the $P(l_o)$ distribution shows a behaviour 
well described by a truncated Pareto distribution. 
The estimation using the MLE gives an exponent of $1.47\pm0.02$ (see Figure \ref{fig_empiric}). 

For performing the analysis of the angles at the turning points,
we measure the net angular change in orientation between two 
consecutive directions along the animal path ($\theta$).
The corresponding $P(\theta)$ distribution can be well described by a 
Gaussian centred around 
zero and with a standard deviation of $58\,^{\circ}$ (see Fig. \ref{fig_empiric}).
An analogous 
behaviour was detected for all the 
species, for oriented and non-oriented individuals
(see Supplementary Information, Figure S1).\\

\begin{figure}[h]
\centering
\includegraphics[width=0.75\textwidth, angle=0]{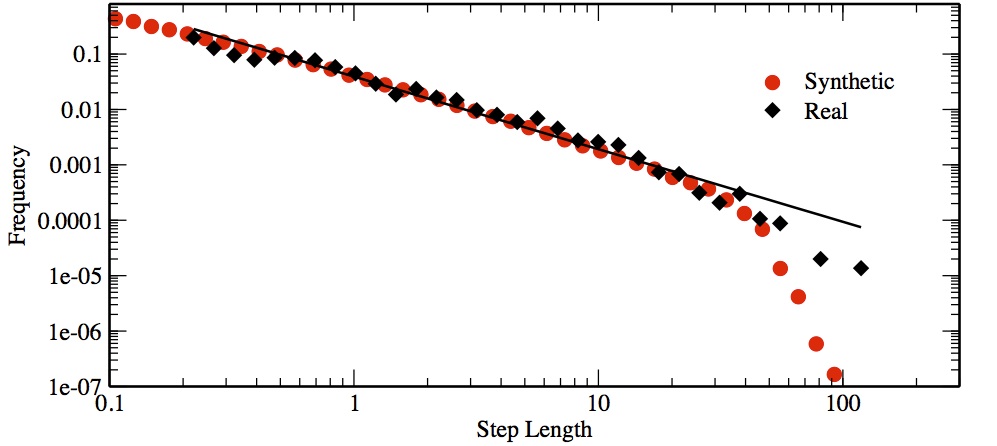}
\caption{{\bf Simulated data.}
Comparison of the histogram of the projected step-lengths
as obtained from the field data (black diamonds) and from the synthetic data
(red dots) generated by our simulation.
These last data show a clear exponential cutoff for large step sizes.
This fact can be ascribed to the process of projection, 
as it is present also for simulations with Pareto truncated distribution
with no correlations in the angles.
The continuous line represents the best
estimated Pareto-truncated distribution for the synthetic data.
 }
\label{fig_simul}
\end{figure}

Given the distribution $P(l_o)$ and $P(\theta)$, 
we are able to build up the simulation for synthetically reproducing the
empirical data-set.
This is obtained performing a discrete walk which implements,
at the same time, both a Pareto truncated step-length distribution and 
a Normal distribution for the turning angles.
We ran a long simulation generating a walk of $5\times10^4$ steps.
For this simulation we fixed the maximum value of the step-length
to $48$ m, and the minimal value equal to $0.01$ m. This corresponds
 to consider an upper cutoff, as effectively measured from the field data,
and a minimal one, to account for the necessary mathematical constraint
of the distribution. 
The same analysis performed on the original field data was applied
to this synthetic data-set: the simulated path is projected on the $x$ and $y$ axis 
to obtain 
the corresponding step-lengths for the projected data. 
As can be observed in Figure \ref{fig_simul}, the synthetic projected data perfectly reproduce 
the behaviour of the real projected data, presenting an estimated $\mu$ value 
equal to $1.299\pm0.003$.

\begin{figure}[h]
\centering
\includegraphics[width=0.75\textwidth, angle=0]{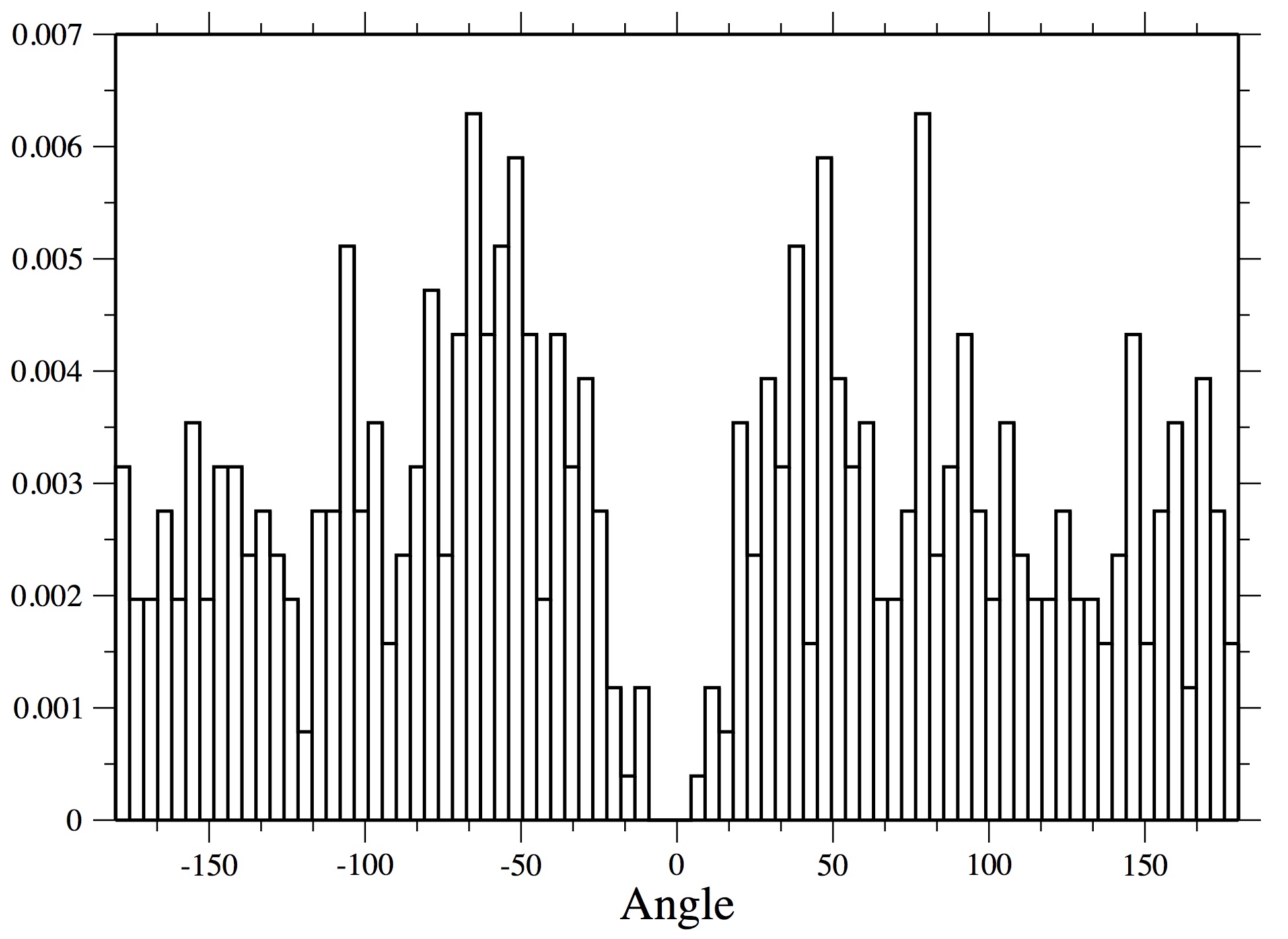}
\caption{{\bf Turning angles.}
Distribution of the turning angles for oriented {\it Philander frenatus}
when measured in correspondence of the points identified as turning points in one dimension, combining the angles in the x and y dimension.
As can be appreciated, 
angles sufficiently far from zero are distributed approximately uniformly.
The behaviour around zero is produced  by the effect of coalescence of small turning angles during the projection operation and by the fact that the recorded changes of direction are always greater than five degrees.
Similar results can be obtained for all the considered dataset.
 }
\label{fig:rangle}
\end{figure}

\section*{Discussion}


Small scale movements of the the three species of marsupials are best described by L\'evy flights. The projection method, $MLE$ and $AIC$ model selection give strong support to L\'evy flights as the best model compared to alternatives. 
The possible existence of a marked scale in terms of an exponential truncation is ruled out by the results of the sequential maximum likelihood estimation on different subsets of our data. The marginal deterioration
of the support in favour of the truncated Pareto distribution 
can be interpreted considering the presence of measurement truncation effects and the use of the projection method for truncated Pareto distribution, as clearly shown in Figure \ref{fig_simul}.

These conclusions were reinforced by the outputs of
the original data-set and the simulated data.
The original data-set 
present a step distribution
well described by a truncated Pareto distribution and
a turning angle $P(\theta)$ distribution well approximated by a 
Gaussian one, similarly to the observed in the turning angles of 
wandering albatrosses \cite{23}.
This result is consonant with the fact that
the motion of these animals presents
a general directional persistence, which causes small turns centred 
around the previous direction. 
It is important to note that a genuine L\'evy flight is defined 
with a $P(\theta)$ given by a random uniform distribution.
However, 
we 
care for being able to describe the projected data as
a standard L\'evy flight and not the original empirical data,
which present an arbitrary segmentation.
In the case of projected data, the turns must be calculated only when 
they correspond to turning points (inversions) in one dimension.
In this case, a uniform distribution was obtained, as can be observed 
in Figure \ref{fig:rangle}. 
Moreover, this projecting operation results in the coalescence of small angles, 
which coherently generates longer step-length, and it accounts for the directional 
persistence of the paths. For this reason, the $\mu$ value for the 
distribution of $P(l_o)$ is larger than the one obtained for the projected data.
The results of the simulation 
corroborates these considerations:
given a walk with a truncated Pareto distribution and 
a strong correlation in the direction of motion, the distribution of the 
projected steps 
presents a clear Pareto truncated distribution with a smaller exponent. 
In particular, the value of the exponent 
obtained with our simulation
is coherent with the one extracted from the real data.
\\
 

Our study is developed on spatial and temporal scales smaller than
the ones usually considered in previous experiments with 
vertebrates \cite{57,92,94}.
In those experiments, animals were generally medium and large sized and they were 
treading long distances, with data usually corresponding to a discrete 
sample of the trajectory.
In contrast, our study was realised with small vertebrates, 
in an area smaller than $0.25$ $km^2$,
using a continuous tracking method which allows an 
accurate description of the trajectories shape.
The fact that also in this condition L\'evy flights are detected,
reinforces the idea that these phenomena are characterized by
a scale-free process, where the same behaviour can be
recorded at different scales \cite{30,42}.

The characterisation of the movement based on a L\'evy flight 
is common to the three species of marsupials, even if they present different 
habits. 
In fact, {\it M. paraguaiana} is a more arboreal species, {\it P. frenatus} is generally terrestrial, 
and {\it D. aurita} presents both arboreal and terrestrial habits \cite{84}.
This result suggests that these patterns of movement present an interspecific 
and perhaps evolutionary character \cite{42,wosniack}, which makes it recurrent between animals of the same family.
L\'evy patterns of movement have been reported for other 
taxonomic groups of mammals, such as primates \cite{92}, 
and ungulates \cite{94,95}, but
this is the first analysis for marsupials.

A rigorous comparison of the $\mu$ values for all the 
different classes of data should be taken with some caution.
In fact, even if the estimated standard error 
can seem relative small, 
this is only the error associated with the fitting,
while other sources of uncertainty,
originated by the data collection, 
the projecting procedure and the data selection, 
were not accounted for.
These considerations are valid particularly
for {\it M. paraguayana}, where few observations
are available (see table \ref{table}) and the final sample 
size is relatively small, a fact that suggests prudence.
In fact, as the  behaviour of the trajectories  is generally 
heterogeneous, the average over different 
observations is important. 

Taking into account these considerations,
for the more robust case of {\it Philander}
we can observe a slight decrease in the exponent 
value for non-oriented compared to oriented animals.
This fact is consonant with
more frequent long steps for oriented animals.
Even so,  
the difference in the $\mu$ values is not 
significative enough to suggest 
a really different strategy of movement
between oriented and non-oriented animals. 

The $\mu$ values of both oriented and non-oriented individuals
correspond to enhanced super-diffusion, approaching the ballistic limit, 
with no tendency towards random behaviours for non-oriented
animals. 
The $\mu$
value is smaller than 1.4 for all the data-set, 
approaching the ballistic limit for the smaller 
exponents.
Such small values have been previously identified in empirical 
studies for foraging albatrosses ($\mu=1.75$ for the average data set, but with individuals values as low as $1.14$) \cite{23}, for jellyfish  ($\mu = 1.18$) \cite{95b} and 
marine predators ($\mu = 1.63$) \cite{32}.

Some of our estimated $\mu$ 
values are relatively close to the value 
$\mu=4/3$, which is found in the partial power-law behaviour present in 1-dimensional Continuous Correlated Random Walks, CCRW \cite{20,48}. 
However, in this case, a clear
and relevant 
exponential truncation must be present,
which did not happen
in our study. 
Also, the 95\% confidence intervals for our $\mu$ 
values do not include the expected $4/3$ value (see Table 2),
but note that these estimations do not account for exponential truncation.
The only exception would be the oriented {\it Philander frenatus},
($\mu= 1.31\pm 0.02$), but again 
only a very feeble exponential truncation
can be suggested (see Figure 2 and 3).
Our results could only be consistent with correlated random walks if the small arena precluded the detection of the real form of the exponential truncation, an aspect which should be object 
of future studies.
Previous empirical studies exploring the connection between L\'evy patterns and CCRW
can be found in \cite{20} for the case of a small arthropod.
Johnson \cite{johnson} reported that autocorrelation timescales of harbor seal and northern fur seal are several hours long, but L\'evy walk movement patterns on these scales have not been measured. L\'evy walk movement patterns with $\mu= 1.25$ can be found in gray seals \cite{17}.\\


The differences in the exponents of the truncated power-law models
recorded in the two considered classes of behaviours are 
relatively weak.
In contrast, a marked and relevant difference
can be observed in the upper truncation value ($b$) of the
L\'evy walks, which for oriented animals
is  roughly twice the size of the not
oriented case (see table \ref{table}).
This fact suggests that the difference in the movement strategy
between the two behaviours can be found not in adjustments in 
the functional form that controls their walks, 
but in the length steps between turning points,
which are in general considerably longer
in the case of oriented animals.
Therefore, animals could be
adjusting the truncation scale of step lengths to switch between oriented and non-oriented 
movements.
This phenomenon can be seen clearer 
looking at the mean number of reorientation
points of a single walk. 
In the case of oriented animals,
the number of reorientation
points is 
roughly half that for not oriented animals.
This is 
estimated considering that 
the number of reorientation
points corresponds to the mean sample size 
of a single observation, in the approximation
that all animals uses practically
the entire length of the spool.
From this spartan estimation, we obtain
around 10 turning points for oriented and 
20 for not oriented animals (see table \ref{table}).\\

Le\'vy flights are considered to be the best solution
for solving the problem of random search 
for sparsely distributed targets \cite{16b,30}.
For this reason, the vast majority of empirical studies that reports
L\'evy processes are observations of foraging behaviour \cite{16,20,32}.
In contrast, the present study considers movements
of animals that had been translocated to an unfamiliar site and 
different habitat \cite{74} (open areas in the matrix). 
Translocated animals generally try to return to familiar habitat and environment upon release, without foraging \cite{stamps}. 
Forested areas are safe and familiar habitat for the three marsupials, while open areas represent 
unfavourable habitats for these animals where the 
predation risk 
by raptors, snakes and domestic dogs is high \cite{10}.  
We note that  even in the absence of predation events
in the course of the experiments, animals may still show a behaviour typical of avoiding predation \cite{12,107}. 
In fact, an unfamiliar environment induces animals to keep alert of potential 
predators and to find a safe place as soon as possible. 
Risk of predation generates a permanent stress, 
and takes a relevant part on the decision-making process \cite{106}. 
For this reason, our empirical results seem to verify
simulations of survival in patchy landscapes, which
shows a generally decrease with $\mu$ \cite{Niebuhr}.
L\'evy walks are described 
outside of the context of optimal foraging only in a few cases,
such as for 
shearwaters 
searching for their breeding colony after crossing vast regions of open ocean \cite{pelagic}
and for bacteria \cite{Ariel}.
\\

Finally, an important difference from previous studies, is that, 
in our case, animals have no previous knowledge of the area of movement.
For this reason, the detected movements 
are not the consequence of memory or of the familiarity with the specific release site.  
It is not plausible to completely rule out the possibility that the detected L\'evy flights are the result of innate behaviour \cite{64}, rather than an optimal strategy to search for safe patches under predation risk. 
The fact that L\'evy patterns are related to some general innate behaviour,
beyond the confines of optimal foraging,
has gain interest in the literature \cite{25}. 
The analysis of movements of {\it P. frenatus} released in similar conditions, but inside forest fragments, presented $\mu$ values comparable to the values estimated in the matrix ( $\mu= 1.37\pm0.02$) (unpubl. data), supporting L\'evy flights as result of innate behaviour.
\\

To sum up, we describe by means of 
L\'evy flights small-scale movements not associated with foraging,
but with animals looking for a shelter in a risky habitat, 
where the memory 
of the area of movement can not play any role.
The result is not dependent on the differences in the habits of the three species,
and a weak differentiation between oriented and non-oriented animals
is reported in the functional form of the truncated power-law behaviour.




\section*{Acknowledgements}

It is with great sadness that, after the submission of this manuscript, 
we must report the passing of the first author and dear colleague Boris Rios Uzeda,
an enthusiastic Bolivian biologist who studied 
the ecology and conservation of mammals in the Atlantic forest, in the Pantanal of Brazil,
and in the Andean region of Bolivia.
We will miss the important contributions he made to our group, where his passion and high standards contributed greatly to the quality of our research, and his work for the conservation 
of wildlife in Latin America. 
Even more, we will miss him as a dear friend.\\

We thank students for assistance in the fieldwork, and staff of the Laborat\'orio de Vertebrados (Universidade Federal do Rio de Janeiro), particularly Angela Marcondes and N\'elio Barros. 
Financial support was provided by grants from PROBIO II/MCTI/MMA/GEF, CNPq PPBio/Rede BioM.A. (457522/2012-7), CNPq PELD (403840/2012-0), FAPERJ CNE (201344/2014), CNPq Universal (461852/2014-4), and CNPq Produtividade em Pesquisa (308974/2015-8). Boris R\'{\i}os-Uzeda was supported by a scholarship from CAPES PEC-PG.

\section*{Author contributions statement}
B.R.U., E.B., and M.V.V. conceived the paper; data were collected by the team of B.R.U. and M.V.V.; E.B. and B.R.U. analysed the data. 
All authors wrote and reviewed the manuscript.

\section*{Additional Information}
\textbf{Competing interests}: The authors declare no competing interests.

\end{document}